\documentclass[11pt]{article}
\usepackage{graphicx,cite,amssymb,epsfig,float,psfrag,multirow,amsmath,subfigure}
\def\Q{{\cal O}}

\newcommand{\noi}{\noindent}

\begin{document}

\thispagestyle{empty} 
\begin{flushright}
\begin{tabular}{l}
{\tt DAPNIA-03-73}\\
{\tt LPT Orsay-03-29 }\\
{\tt LMU-03-08}\\
{\tt PCCF-RI-0305}\\
{\tt hep-ph/0304205}\\
\end{tabular}
\end{flushright}
\vskip 2.2cm\par
\begin{center}
{\par\centering \textbf{\LARGE Phenomenological discussion of $B\to P V$ decays\\
in QCD improved factorization approach}\footnote{Talk presented at the XXXVIIIth Rencontres de Moriond: Electroweak Interactions and Unified Theories, Les Arcs, France, March 15-22, 2003.} }
\\


\vskip 0.9cm\par
{\par\centering \large  

R.~Aleksan$^a$,
P.-F.~Giraud$^a$,
V.~Mor\'enas$^b$,
O.~P\`ene$^c$ and 
\underline{A. S. Safir}$^d$.
}
{\par\centering \vskip 0.5 cm\par}
{\par\centering \textsl{ 
$^a$~DSM/DAPNIA/SPP, CEA/Saclay, 
F-91191 Gif-sur-Yvette Cedex, France.
\\
$^b$~LPC, Universit\'e Blaise Pascal - CNRS/IN2P3 F-63000 
Aubi{\`e}re Cedex, France. \\
$^c$~LPT (B\^at.210), Universit\'e de Paris XI, Centre d'Orsay, 91405
Orsay-Cedex, France. \\  
$^d$~LMU M\"unchen, Sektion Physik, Theresienstra\ss e 37, D-80333
M\"unchen, Germany.} \\ 
\vskip 0.2cm\par}
\today
 \end{center}

\vspace{1cm}
\begin{abstract}
Trying a global fit  of the experimental branching ratios  and CP-asymmetries
of  the charmless $B\to PV$  decays according to QCD factorization, we find it impossible to reach  a satisfactory
 agreement, the confidence level (CL) of the best fit is smaller than .1 \%.
  This  failure reflects the difficulty
 to accommodate several large  experimental branching ratios of the strange
 channels. Furthermore, experiment was not able to exclude a large direct CP
asymmetry in $\overline {B}^0\to\rho^+ \pi^-$, which is predicted very small
by QCD factorization. Proposing a fit with QCD factorization 
 complemented  by a charming-penguin inspired model we reach a best fit 
 which is not excluded by experiment (CL of about 8 \%) but is not fully convincing.
  These negative results must be tempered  by the remark that some of the
  experimental data used are recent and might still  evolve significantly. 
\end{abstract}
\vspace{1cm}
\section{Introduction}
\hspace*{\parindent}
It is well known that the non-leptonic decay and particularly the non-leptonic
 $B$ decay is one of the most exciting and challenging sector in the Standard Model, especially due to its non-perturbative regime. A good understanding of these transitions will not only provide a good estimate of the CKM parameters and the CP violating parameters (particularly the so-called angle $\alpha$ of the unitarity triangle\footnote{It is well known that extracting $\alpha$ from measured indirect CP asymmetries needs a sufficient control of the relative size of the so-called tree ($T$) and penguins ($P$) amplitudes.}), but also of the hadronic dynamics, such as form factors and long distances contributions.

Experimentally, many branching ratios and CP asymmetries of two body non-leptonic B-decays, especially charmless $B\to P V$ decays, have been reported recently with increasing accuracy by BaBar, Belle and CLEO (see \cite{hep-ph/0301165} and references therein), providing a crucial challenge to the theory, which is a difficult issue in this sector.
 
Since long, one has used what  is  now called ``naive factorization'' which 
replaces the matrix element of a four-fermion operator in a heavy-quark 
decay by the product of the matrix elements of two currents, one semi-leptonic matrix element and one purely leptonic. 
For long it was noticed that  naive factorization did provide reasonable 
results although it was impossible to derive it rigorously
from QCD except in the $N_c \to  \infty$ limit. It was also well-known
that the matrix elements computed via naive factorization 
have a wrong anomalous dimension.
Recently an important theoretical progress has been performed~\cite{QCDF,0104110}
which is commonly called ``QCD factorization''. It is based on the fact that 
the $b$ quark is heavy compared to the intrinsic scale of strong 
interactions. This allows  to deduce that non-leptonic decay amplitudes 
in the heavy-quark limit have a simple structure. It implies 
that corrections termed ``non-factorizable'', 
which were thought to be intractable, can be calculated rigorously. The anomalous dimension of the matrix elements is now correct to the order
at which the calculation is performed. 
Unluckily the subleading ${\cal  O}(\Lambda/m_b)$ contributions cannot in general be computed rigorously because of infrared singularities, and some 
competitives chirally enhanced terms, which could justify
 a significantly larger bound  with the risk of seeing these unpredictable 
 terms become dominant~\cite{QCDF,0104110}.  
It is then  of utmost importance to check experimentally  QCD factorization.

Since a few years it has been applied to $B\to PP$ (two charmless pseudoscalar
 mesons)
decays. The general feature is that the decay to non-strange final states 
is predicted slightly larger than experiment while the decay to strange
final states is significantly underestimated. In~\cite{0104110} it is claimed that
this can be cured by a value of the unitarity-triangle angle $\gamma$ larger
than generally expected, larger maybe than 90 degrees. Taking also into account
various uncertainties the authors conclude positively as for the agreement of 
QCD factorization with the data. In~\cite{Isola:2001bn,Ciuchini:2001gv} it was objected that the large branching ratios
for strange channels argued in favor of the presence of a specific non
perturbative contribution called ``charming
penguins''~\cite{Ciuchini:2001gv,Chiang:2000eh,Ciuchini:1997rj,Ciuchini:1999yv,Ciuchini:1997hb}. We will return to
this approach later.

The  $B\to PV$ (charmless pseudoscalar + vector mesons) channels are
 more numerous and allow a more extensive check. In ref. \cite{Aleksan:1995wn} it was shown that 
naive factorization implied a rather small 
$|P|/|T|$ ratio,  for  $\overline {B}^0\to\rho^\pm\pi^\mp$ decay 
channel, to be
compared to the larger one for the  $B\to\pi^+\pi^-$. 
This prediction is still valid in QCD factorization where 
the $|P|/|T|$ ratio is of about 3 \% (8 \%) for the 
$\overline {B}^0\to\rho^+\pi^-$ ($\overline {B}^0\to\rho^-\pi^+$)  channel
against about 20 \% for the $\overline {B}^0\to\pi^+\pi^-$ one.
If this
prediction was reliable it would put the $\overline {B}^0\to\rho^+\pi^-$ channel
in a  good position to measure the CKM angle $\alpha$ via indirect
CP violation. This remark triggered the present work: we wanted to 
check QCD factorization in the $B\to PV$ sector to  estimate the chances 
for a relatively easy determination of the angle $\alpha$.

 The non-charmed $B\to PV$  amplitudes have been 
computed in naive factorization~\cite{ali}, in some extension of 
naive factorization including strong phases~\cite{Kramer:1994in},
in QCD
factorization~\cite{Yang:2000xn,Du:2002up,Du:2002cf} and some of them in the
so-called perturbative
QCD~\cite{Chen:2001pr}. In~\cite{Cottingham:2001ax}, a 
global fit to $B\to PP,PV,VV$ was investigated using QCDF in 
the heavy quark limit  and it has been found a
plausible set of soft QCD parameters that apart from three pseudoscalar
vector channels, fit well the experimental branching ratios.
Recently~in \cite{Du:2002cf} it was
claimed from a global fit to $B\to PP,PV$ that the
predictions of QCD factorization are in good agreement with 
experiment when one excludes the channels containing a $K^*$ in the final states, from the global fit. 

In this talk, we present the result of \cite{hep-ph/0301165}, where a systematic analysis of the charmless $B\to P~V$ decays was performed in order to understand and try to settle the present status of the comparison of QCD factorization with experiment.

\section{Theoretical framework}
\label{sec:Heff}
\begin{figure}
\epsfxsize=10cm
\centerline{\epsffile{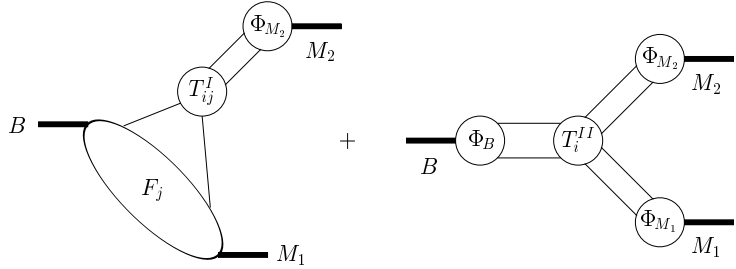}}
\centerline{ \parbox{16cm}
 {\caption{ \label{fig0}\it
Graphical representation of the factorization formula (\protect\ref{fact}), with only one of the two form-factor shown.
}}}
\end{figure}

\hspace*{\parindent}
When the  QCD factorization (QCDF) method is applied to the decays
$B{\to}PV$, the hadronic matrix elements of the local effective operators can be written as

\begin{eqnarray}\label{fact}
   \langle\ P\, V|\Q_i|B\rangle
   &=& F_1^{B\to P}(0)\,T_{V,i}^{\rm I}\star f_V\Phi_V
    + A_0^{B\to V}(0)\,T_{P,i}^{\rm I}\star f_P\Phi_P \nonumber\\
   &&\mbox{}+ T_i^{\rm II} \star f_B\Phi_B \star f_V\Phi_V \star f_P\Phi_P \,,
\end{eqnarray}
where $\Phi_M$ are leading-twist light-cone distribution amplitudes, 
and the $\star$-products imply an integration over the light-cone momentum 
fractions of the constituent quarks inside the mesons. A graphical 
representation of this result is shown in figure \ref{fig0}.

 Here $F_1^{B{\to}P}$ and $A_0^{B{\to}V}$ denote the form factors for $B{\to}P$
 and $B{\to}V$ transitions, respectively. ${\Phi}_{B}(\xi)$,
 ${\Phi}_{V}(x)$, and ${\Phi}_{P}(y)$ are the light-cone distribution
 amplitudes (LCDA) of valence quark Fock
 states for $B$, vector, and pseudoscalar mesons, respectively.
 $T^{I,II}_{i}$ denote the hard-scattering kernels, which are dominated by
 hard gluon exchange when the power suppressed 
 ${\cal O}({\Lambda}_{QCD}/m_{b})$ terms are neglected. So they are
 calculable order by order in perturbation theory. 
One of the most interesting results of the QCDF approach is that, in the 
heavy quark limit, the strong phases arise naturally from the hard-scattering kernels
 at the order of ${\alpha}_{s}$. 
 As for the nonperturbative part, they are, as already mentioned, taken 
 into account by the form factors and the LCDA of mesons up to corrections
 which are power suppressed in $1/m_b$.

In QCDF 
, the decay amplitudes for $B{\to}P\,\,V$ in the
 heavy quark limit can be written as
 \begin{equation}
 {\cal A}(B{\to}PV) \,\, {\propto}\,\, 
    \sum\limits_{p=u,c} \sum\limits_{i=1}^{10} \lambda_{p}\,\, a_{i}^{p}\,\,
   {\langle}P\,\, V{\vert}{\cal O}_{i}{\vert}B{\rangle}_{nf}.
 \label{eq:decay-f}
 \end{equation}
 The above ${\langle}PV{\vert}{\cal O}_{i}{\vert}B{\rangle}_{ nf}$ are the 
 factorized hadronic matrix elements, which have the same definitions as
 those in the NF approach. The ``nonfactorizable'' effects are included in
 the coefficients $a_{i}$ which are process dependent. The coefficients
 $a_{i}$ and the explicit expressions for the decay amplitudes of $B{\to}P\,\,V$ can be found in ref.~\cite{hep-ph/0301165}. 

 According to the arguments in \cite{QCDF}, the contributions of weak
 annihilation to the decay amplitudes are power suppressed, and they
 do not appear in the QCDF formula (\ref{fact}). But, as emphasized
 in \cite{Nagashima:2002ia,0004173,0004213}, the contributions from weak annihilation
 could give large strong phases with QCD corrections, and hence large CP
 violation could be expected, so their effects cannot simply be neglected.
 However, in the QCDF method, the annihilation topologies
 violate factorization because of the endpoint divergence, which could be unfortunately controlled just as a phenomenological parameters~\cite{0104110}. 
In this work, we will follow the treatment of ref.
 \cite{0104110} and express the weak annihilation topological decay
 amplitudes as
 \begin{equation}
 {\cal A}^{a}(B{\to}PV)\,\, {\propto}\,\, f_{B}\,\, f_{P}\,\, f_{V}
    {\sum} \lambda_{p}\,\, b_{i} \; ,
 \label{eq:decay-a}
 \end{equation}
 where the parameters $b_{i}$ are collected in~\cite{hep-ph/0301165}.
\section{ QCD factorisation  versus experiment}
\label{exp}
\hspace*{\parindent}
In order to propose a test of QCD factorization with respect to
experiment, a compilation of various charmless branching fractions and
direct $CP$ asymmetries was performed. This
compilation includes the latest results from BaBar, Belle and CLEO. 
\begin{figure}
  \psfrag{chi2}[tr][tr]{\large $\chi^2$}
  \psfrag{experiments}[br][br]{\large experiments}
  \begin{center}
  \subfigure[Scenario 1]{\includegraphics[width=6cm]{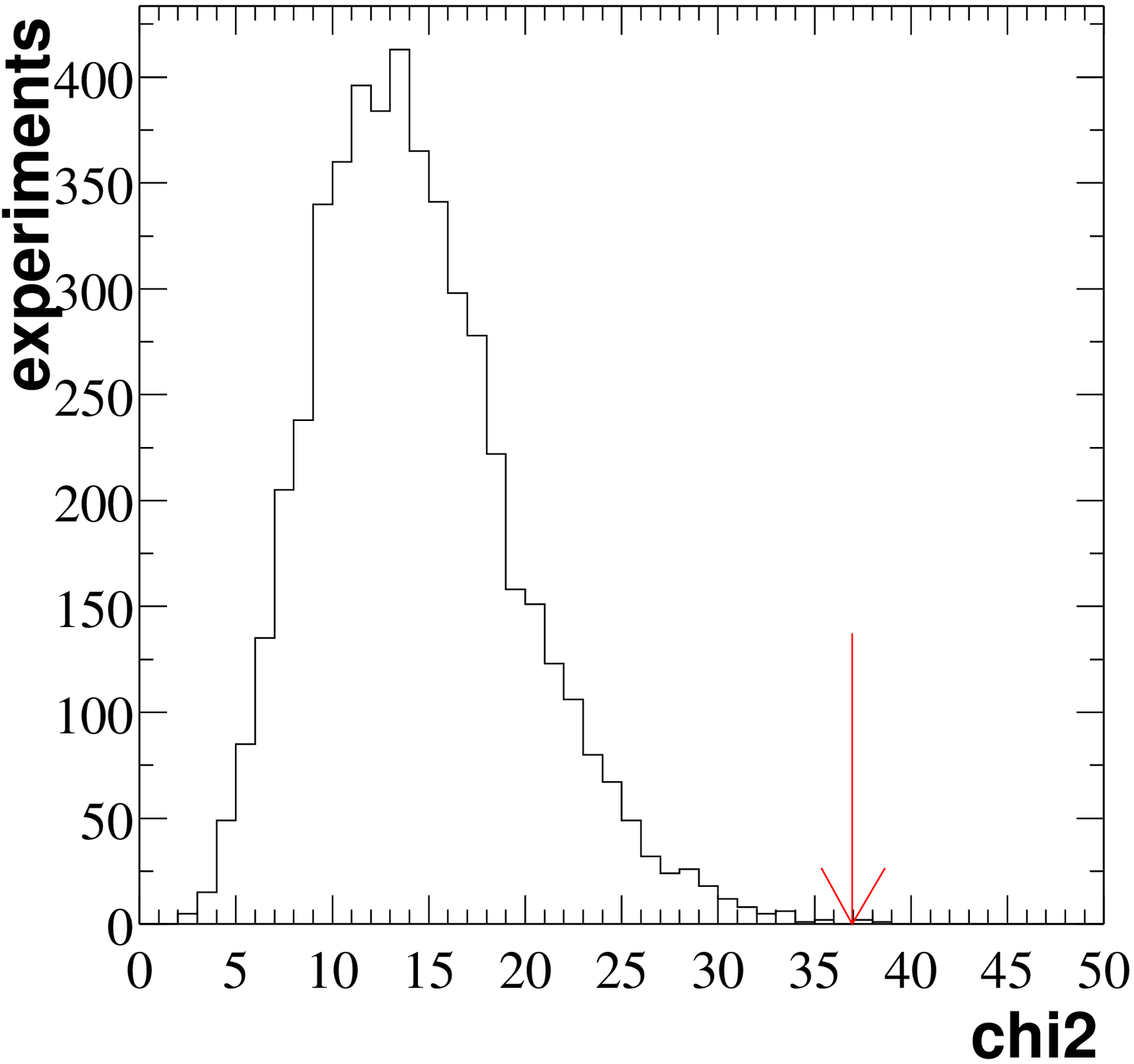}}
  \hspace{5mm}
  \subfigure[Scenario 2]{\includegraphics[width=6cm]{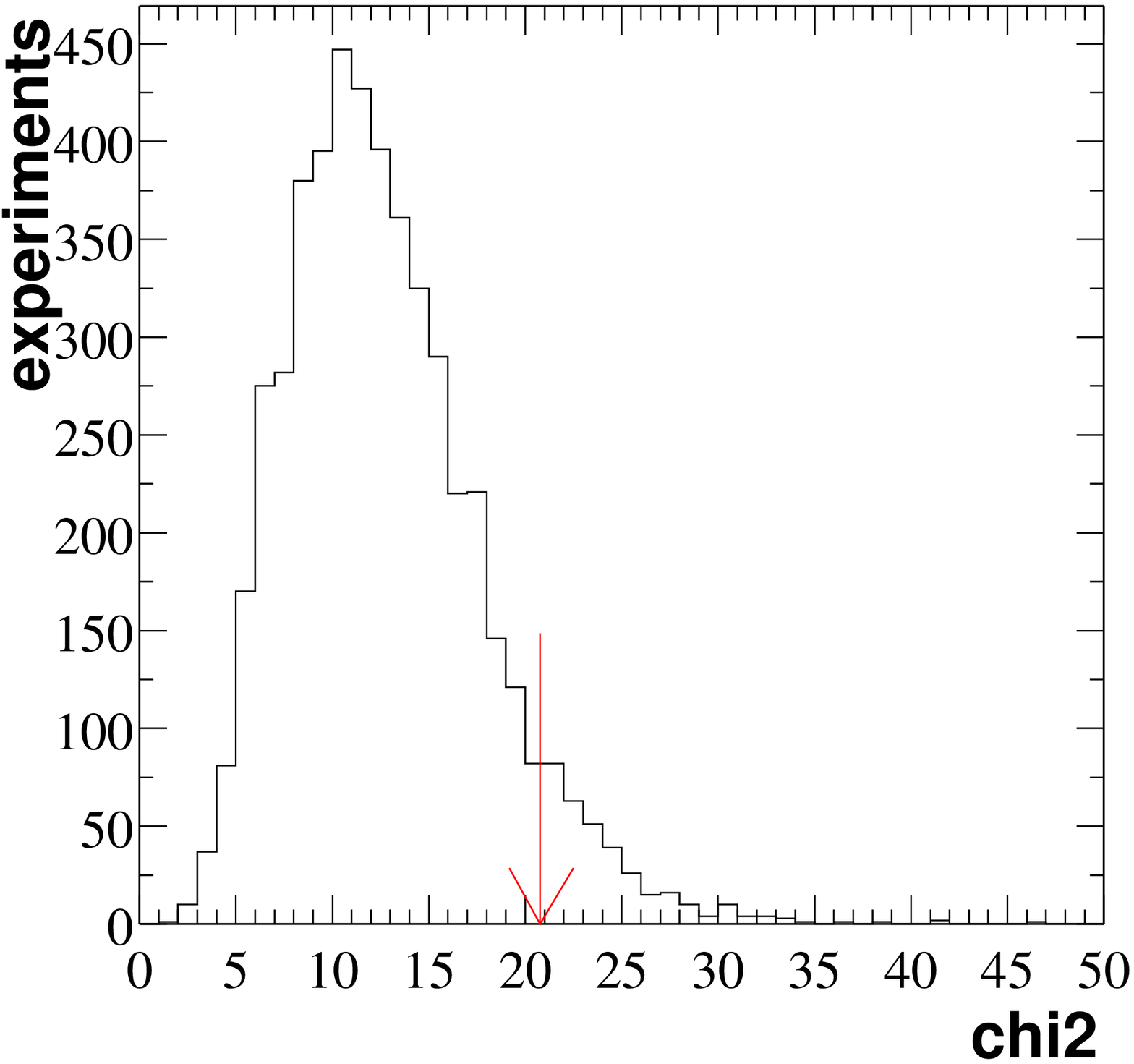}}
  \end{center}
  {\caption{\it Goodness of fit test of the two proposed scenarios: 
      the arrow points at the value $\chi^2_{\mathrm{data}}$
    found from the measurements, and the histogram shows the corresponding 
    $\chi^2$ in the case that the models predictions are
    correct.}}
  \label{fig:goodness}
\end{figure}

In order to compare the theoretical predictions with the
experimental measurements, we have computed the $\chi^2$
and then  minimized it using MINUIT~\cite{MINUIT}, letting
free all theoretical parameters in their allowed range. 
The theoretical predictions,  with the theoretical parameters yielding 
the best fits, are compared to
experiment in table \ref{tab:fit} for two scenarios to be explained below. 
 The asymmetries of the $\rho^\pm\pi^\mp$  channels can be 
 expressed (see ref.~[9] in \cite{hep-ph/0301165})
 in terms of the quantities reported in table 4 of \cite{hep-ph/0301165}. The comparison 
 between their theoretical predictions and experiment is reported in  
  table~\ref{tab:asym}.
 
Scenario 1 refers to a fit according to QCD factorization,
varying all theoretical parameters in the range presented in table
\ref{tab:input}.  Even the unitarity triangle angle $\gamma$ is varied
freely and ends up not far from 90 degrees. To label our ignorance of the non perturbatively calculable subdominant contribution to the annihilation and hard
scattering, we have taken for simplicity $X_A=X_H$
in the range proposed in ref.~\cite{0104110}:
\begin{equation}\label{eq:xa}
X_{A,H}=\int \limits_0^1 \frac{dx}{x} = \ln \frac{m_B}{\Lambda_{h}}
(1+\rho_{A,H}\,e^{i\phi_{A,H}}).
\end{equation}

Scenario 2 in table \ref{tab:fit}  refers to  a fit adding  a charming penguin
inspired long distance contribution which will be presented and discussed in  
section~\ref{long}. In this fit $\gamma$ is constrained within the range
$[34^\circ, 82^\circ]$. 

The values of the theoretical parameters found for the two best fits
is given in table \ref{tab:input}: many parameters are found to be at
the edge of their allowed range\footnote{In Table 1, the fit 
value of $\rho_A$ appears at the edge of the input range, $\rho_A=1$. 
However enlarging its range, such  as $|\rho_A|\leq 10$, 
brings a large annihilation contributions  $\{\rho_A,\phi_A\}= \{2.3(4.4), -41(-108)^{o}\}$ for scenario 1 (2).
}. 
In order to estimate the quality of
the agreement between measurements and predictions, the standard
Monte Carlo based ``goodness of fit'' test was performed (see ref.~\cite{hep-ph/0301165} for further details).
The results of the ``goodness of fit'' tests are given in
figure~2, where, one obtains an
upper limit for the confidence level in scenario 1, $CL \le 0.1\%$ and $CL \le 7.7\%$ for the scenario 2.
\renewcommand{\arraystretch}{1.3}
\begin{table}[t]
\begin{center}
  \begin{tabular}{|l|c|c|c|}\hline
Input & Range & Scenario 1 & Scenario 2 \\ \hline
$\gamma$ (deg)&  & $99.955$ & $81.933$\\
$m_s$ (GeV)& $[0.085,0.135]$ & $0.085$ & $0.085$\\
$\mu$ (GeV)& $[2.1,8.4]$ & $3.355$ & $5.971$\\
$\rho_A$& $[-1,1]$ & $1.000$ & $1.000$\\
$\phi_A$(deg)& $[-180,180]$ & $-22.928$ & $-87.907$\\
$\lambda_B$ (GeV)& $[0.2,0.5]$ & $0.500$ & $0.500$\\
$f_B$   (GeV)& $[0.14,0.22]$ & $0.220$ & $0.203$\\
$R_u$& $[0.35,0.49]$ & $0.350$ & $0.350$\\
$R_c$& $[0.018,0.025]$ & $0.018$ & $0.018$\\
$A_0^{B\to \rho}$& $[0.3162,0.4278]$ & $0.373$ & $0.377$\\
$F_1^{B\to \pi}$& $[0.23,0.33]$ & $0.330$ & $0.301$\\
$A_0^{B\to \omega}$& $[0.25,0.35]$ & $0.350$ & $0.326$\\
$A_0^{B\to K^*}$& $[0.3995,0.5405]$ & $0.400$ & $0.469$\\
$F_1^{B\to K}$& $[0.28,0.4]$ & $0.333$ & $0.280$\\
${\mathrm Re}[{\mathcal A}^{P}]$& $[-0.01,0.01]$ & & $0.00253$\\
${\mathrm Im}[{\mathcal A}^{P}]$& $[-0.01,0.01]$ & & $-0.00181$\\
${\mathrm Re}[{\mathcal A}^{V}]$& $[-0.01,0.01]$ & & $-0.00187$\\
${\mathrm Im}[{\mathcal A}^{V}]$& $[-0.01,0.01]$ & & $0.00049$\\\hline
  \end{tabular}
\end{center}
\centerline{\parbox{16cm}
  {\caption{\label{tab:input}\it Various theoretical inputs used in our global
  analysis of $B\,\to\,PV$  decays in QCDF. The parameter ranges have 
  been taken from literature~\cite{0104110,Yang:2000xn,Du:2002up,Ball:1998kk}.
  The two last columns give the best fits of both scenarios.}}}
\end{table}
\renewcommand{\arraystretch}{1.3}
\begin{table}
\begin{center}
  \begin{tabular}{|l|c|cc|cc|}
\hline
 & Experiment & \multicolumn{2}{c|}{Scenario 1} & \multicolumn{2}{c|}{Scenario 2} \\ 
 & & Prediction & $\chi^2$ & Prediction & $\chi^2$ \\ \hline
${\mathcal BR}({\overline B}^0\,\to\,\rho^0\,\pi^0)$ & $2.07 \pm 1.88$ & 0.132 & 1.1 & 0.177 & 1.0\\
${\mathcal BR}({\overline B}^0\,\to\,\rho^+\,\pi^-)$ &  & 11.023 &  & 10.962 & \\
${\mathcal BR}({\overline B}^0\,\to\,\rho^-\,\pi^+)$ &  & 18.374 &  & 17.429 & \\
${\mathcal BR}({\overline B}^0\,\to\,\rho^{\pm}\,\pi^{\mp})$ & $25.53 \pm 4.32$ & 29.397 & 0.8 & 28.391 & 0.4\\
${\mathcal BR}(B^-\,\to\,\rho^0\,\pi^-)$ & $9.49 \pm 2.57$ & 9.889 & 0.0 & 7.879 & 0.4\\
${\mathcal BR}(B^-\,\to\,\omega\,\pi^-)$ & $6.22 \pm 1.7$ & 6.002 & 0.0 & 5.186 & 0.4\\
${\mathcal BR}(B^-\,\to\,K^{*-}\,K^0)$ &  & 0.457 &  & 0.788 & \\
${\mathcal BR}(B^-\,\to\,K^{*0}\,K^-)$ &  & 0.490 &  & 0.494 & \\
${\mathcal BR}(B^-\,\to\,\Phi\,\pi^-)$ &  & 0.004 &  & 0.003 & \\
${\mathcal BR}(B^-\,\to\,\rho^-\,\pi^0)$ &  & 9.646 &  & 11.404 & \\
${\mathcal BR}({\overline B}^0\,\to\,\rho^0\,{\overline K}^0)$ &  & 5.865 &  & 8.893 & \\
${\mathcal BR}({\overline B}^0\,\to\,\omega\,{\overline K}^0)$ & $6.34 \pm 1.82$ & 2.318 & 4.9 & 5.606 & 0.2\\
${\mathcal BR}({\overline B}^0\,\to\,\rho^+\,K^-)$ & $15.88 \pm 4.65$ & 6.531 & 4.0 & 14.304 & 0.1\\
${\mathcal BR}({\overline B}^0\,\to\,K^{*-}\,\pi^+)$ & $19.3 \pm 5.2$ & 9.760 & 3.4 & 10.787 & 2.7\\
${\mathcal BR}( B^-\,\to\,K^{*-}\,\pi^0)$ & $7.1 \pm 11.4$ & 7.303 & 0.0 & 8.292 & 0.0\\
${\mathcal BR}({\overline B}^0\,\to\,\Phi\,{\overline K}^0)$ & $8.72 \pm 1.37$ & 8.360 & 0.1 & 8.898 & 0.0\\
${\mathcal BR}(B^-\,\to\,{\overline K}^{*0}\,\pi^-)$ & $12.12 \pm 3.13$ & 7.889 & 1.8 & 11.080 & 0.1\\
${\mathcal BR}(B^-\,\to\,\rho^0\,K^-)$ & $8.92 \pm 3.6$ & 1.882 & 3.8 & 5.655 & 0.8\\
${\mathcal BR}(B^-\,\to\,\rho^{-}\,{\overline K}^0)$ &  & 7.140 &  & 14.006 & \\
${\mathcal BR}(B^-\,\to\,\omega\,K^-)$ & $2.92 \pm 1.94$ & 2.398 & 0.1 & 6.320 & 3.1\\
${\mathcal BR}(B^-\,\to\,\Phi\,K^-)$ & $8.88 \pm 1.24$ & 8.941 & 0.0 & 9.479 & 0.2\\
${\mathcal BR}({\overline B}^0\,\to\,{\overline K}^{*0}\,\eta)$ & $16.41 \pm 3.21$ & 22.807 & 4.0 & 18.968 & 0.6\\
${\mathcal BR}(B^-\,\to\,K^{*-}\,\eta)$ & $25.4 \pm 5.6$ & 17.855 & 1.8 & 15.543 & 3.1\\
$\Delta\,{\mathcal C_{\rho\pi}}$ & $0.38 \pm 0.23$ & 0.250 & \multirow{4}{1.2cm}{$\left. \begin{matrix} \\ \\ \\ \\
\end{matrix}\right\}8.1/4$} & 0.228 & \multirow{4}{1.2cm}{$\left. \begin{matrix} \\ \\ \\ \\
\end{matrix}\right\}3.9/4$}\\
${\mathcal C_{\rho\pi}}$ & $0.45 \pm 0.21$ & 0.019 &  & 0.092 & \\
${\mathcal A_{CP}^{\rho\,\pi}}$ & $-0.22 \pm 0.11$ & -0.015 &  & -0.115 & \\
${\mathcal A_{CP}^{\rho\,K}}$ & $0.19 \pm 0.18$ & 0.060 &  & 0.197 & \\
${\mathcal A_{CP}^{\omega\,\pi^-}}$ & $-0.21 \pm 0.19$ & -0.072 & 0.5 & -0.198 & 0.0\\
${\mathcal A_{CP}^{\omega\,K^-}}$ & $-0.21 \pm 0.28$ & 0.029 & 0.7 & 0.189 & 2.0\\
${\mathcal A_{CP}^{\eta\,K^{*-}}}$ & $-0.05 \pm 0.3$ & -0.138 & 0.1 & -0.217 & 0.3\\
${\mathcal A_{CP}^{\eta\,{\overline K}^{*0}}}$ & $0.17 \pm 0.28$ & -0.186 & 1.6 & -0.158 & 1.4\\
${\mathcal A_{CP}^{\phi\,K^{-}}}$ & $-0.05 \pm 0.2$ & 0.006 & 0.1 & 0.005 & 0.1\\
\cline{4-4}\cline{6-6}
 &  & & 36.9 & & 20.8\\  
\hline
  \end{tabular}
\end{center}
\centerline{\parbox{16cm}{
\caption{\label{tab:fit}
\it Best fit values using the global analysis of $B\,\to\,PV$ 
decays in QCDF with free $\gamma$ (scenario 1) and QCDF+Charming Penguins 
(scenario 2) with constrained $\gamma$.}}}
\end{table}

\begin{table}[htb!]
\begin{center}
\begin{tabular}{|c|c|c|c|}
\hline
  & Experiment & Scenario 1 & Scenario 2\\ \hline
${\mathcal A_{CP}^{\rho^+\,\pi^-}}$ & $-0.82\pm 0.31\pm 0.16$ & $-0.04$ & $-0.23$ \\
\hline
${\mathcal A_{CP}^{\rho^-\,\pi^+}}$ & $-0.11\pm 0.16\pm 0.09$ & $-0.0002$ & $0.04$ \\
\hline
\end{tabular}
\end{center}
\centerline{\parbox{16cm}{
\caption{\label{tab:asym}\it Values of the CP asymmetries for $B\,\to\,\pi\rho$
decays in QCDF (scenario 1) and QCDF+Charming Penguins (scenario 2).
The notations are explained in ref.~[9] of \cite{hep-ph/0301165}.}}}
\end{table}
 
For the sake of definiteness let us remind that the branching ratios for any
 charmless $B$ decays, $B\to P V$ channel,
 in the rest frame of the $B$-meson, is given by
\begin{equation}\label{eq:Br}
{\mathcal BR}(B\to P V)\propto 
|{\cal A}(B\to P V) + {\cal A}^{a}(B\to P V) + {\cal A}^{\rm LD}(B\to P V)|^2.
\end{equation}

\noi The amplitudes ${\cal A}, {\cal A}^{a}$ and ${\cal A}^{\rm LD}$
are defined in appendix A, B of ref.~\cite{hep-ph/0301165} and 
in eq.~(\ref{eq:LDd}) respectively. 
In the case of pure QCD factorization (scenario 1) we take of course
${\cal A}^{\rm LD} = 0$.  

Our negative conclusion about the QCD factorization  fit
of the $B\to PV$ channels is at odds with the conclusion of the authors
of ref. \cite{Du:2002cf}, who have performed a successful fit of 
both $B\to PP$ and $B\to PV$ channels using the same theoretical starting point, and excluding from their fits the $K^\ast$ final state channels\footnote{These channels seemed questionable to them.}.

We have thus made a fit without the channels containing the $K^\ast$,
 and indeed we find as ref.~\cite{Du:2002cf}
that the global agreement between QCD factorization and experiment was
satisfactory. Notice that this fit was done without discarding  
the channels $B^+\to \omega \pi^+(K^+)$ as done by Du {\it et al}.

Notice also that the parameters $C_{\rho\pi}$ and 
the  $A^{\rho\pi}_{CP}$ have been kept in this fit.
The disagreement between QCDF and experiment for these quantities was
not enough to spoil the satisfactory agreement of the global fit
 because the experimental errors are still large on these quantities.

 \section{A simple model for long distance interactions}
 \label{long}
 \hspace*{\parindent}
As seen in table \ref{tab:fit} the failure of our overall fit with 
QCDF can be traced to two main facts.
First, the strange branching ratios are underestimated by QCDF. Second the 
direct CP asymmetries in the non-strange channels might also be underestimated. 
A priori this could be cured if some non-perturbative mechanism 
was contributing to $|P|$. Indeed, first, in the strange channels, $|P|$
 is Cabibbo 
enhanced and such a non-perturbative contribution could increase 
the branching ratios,
and second, increasing $|P|/|T|$ in the non-strange channels with non-small
strong phases could increase significantly the  direct CP asymmetries
as already discussed. 
We have therefore  tried a model for long distance penguin contributions, namely the charming-penguin inspired model, which depends only
 on two fitted complex numbers\footnote{In order to avoid to add too many new parameters which would make the fit void of signification.}. 
  
 Let us start by describing our charming-penguin inspired  
 model for strange final states.
  In the ``charming penguin'' picture the weak decay of
 a $\overline B^0$ ($B^-$) meson through the action of the 
 operator $Q_1^c= (\bar c b)_{V-A} (\bar s c)_{V-A}$ creates an hadronic system containing 
 the quarks $s, \bar d (\bar u), c, \bar c$, for example 
 $\overline D_s^{(\ast)}$ + $D^{(\ast)}$ systems. This system goes to
 long distances, the $ c, \bar c$ eventually annihilate, a pair
 of light quarks are created by non-perturbative strong interaction 
 and one is left with  two light meson\footnote{We leave aside from now on the $\eta'$ which is presumably quite
  special.}. 
  
Assuming  the flavor-$SU(3)$ symmetry and the OZI rule in the decay amplitude, one can express the long distance term  by two universal complex amplitudes 
respectively as ${\cal A^{P}}$  (${\cal A^{V}}$) when the active quark ends up in the Pseudoscalar (Vector) meson,
weighted by a CG coefficient computed simply by the overlap factor (see \cite{hep-ph/0301165} for further details).
 In practice, to the QCDF's decay amplitudes, we add the long
 distance  amplitudes, given by:
 \begin{equation}\label{eq:LDd}
 {\cal A}^{\rm LD}(B\to PV) = \frac{G_F}{\sqrt{2}}m_B^2 \lambda_p
 (Cl^P\,{\cal A}^P  +Cl^V\, {\cal A}^V).
 \end{equation}
 
The fit with long distance penguin contributions is presented in table 
 \ref{tab:fit} under the label ``Scenario 2''. The agreement with experiment is
 improved, it should be so, but not in such a fully convincing manner. The
 goodness of the fit is about 8 \% which implies that this model is not excluded by
 experiment.  However
 a look at table \ref{tab:input} shows that several fitted parameters are still
 stuck at the end of the allowed range of variation. In particular
 $\rho_A = 1$  means that the uncalculable subleading contribution to QCDF 
 is again stretched to its extreme.
   

\section{Conclusion}
\label{conclusion}
\hspace*{\parindent}
We have made a global fit according to QCD factorization 
of published experimental data concerning charmless 
$B\to PV$ decays including CP asymmetries and excluding 
the channels containing the $\eta'$ meson.
 Our conclusion is that it is impossible to reach a good fit. 
As can be seen in the scenario 1 of table \ref{tab:fit}, the reasons
of this failure is that 
the branching ratios for the strange channels are predicted 
significantly smaller than experiment except for the $B\to \phi K$
channels, and in table \ref{tab:asym} 
it can be seen that the direct CP asymmetry of $\overline {B}^0\to \rho^+\pi^-$
is predicted very small while experiment gives it very large but only two sigmas 
from zero. Not only is the ``goodness of the fit''
smaller than .1 \%, but the fitted parameters show a tendency to  evade
the allowed domain of QCD factorization.

Both the small predicted branching ratios of the strange channels
and the small predicted direct CP asymmetries in the non strange channels
could be blamed on too small $P$ amplitudes
with too small ``strong phases'' relatively to the $T$ amplitudes. 
We have therefore tried the addition of two ``charming penguin'' inspired 
long distance complex amplitudes combined, in order to make the model
predictive enough, with exact flavor-$SU(3)$ and OZI rule. 
This fit is better than the pure QCDF one: with a 
goodness of the fit of about 8 \%, the model is not excluded by experiment.
But the parameters show again a 
tendency to reach the limits of the allowed domain and the best fit 
gives rather small value to the long distance contribution.

Altogether, the present situation is unpleasant. QCDF  seems to be unable
 to comply to experiment. QCDF implemented by an 
ad-hoc long distance model is not fully convincing. No clear
hint for the origin of this problem is provided by the total set of
experimental  data.  If this means that the subdominant unpredictable
contributions are larger than expected, the situation will remain stuck until
some new theoretical ideas are found. 

Maybe  however, the coming experimental data will move enough
to resolve, at least partly, this discrepancy.
We would like to insist on the crucial importance of direct CP asymmetries
in non-strange channels. If they confirm the tendency to be large, 
this would make the case for QCDF really difficult. 

 \section*{Acknowledgements}
\hspace*{\parindent}
A.~Salim Safir would like to thank the organizers of this conference for their financial support.


 
\end{document}